\documentstyle[12pt,a4]{article}

\input{tcilatex}
\begin{document}

\setcounter{page}{0} \topmargin 0pt \oddsidemargin 5mm \renewcommand{%
\thefootnote}{\fnsymbol{footnote}} \newpage \setcounter{page}{0} 
\begin{titlepage}
\begin{flushright}
Berlin Sfb288 Preprint  \\
hep-th/9711113\\
\end{flushright}
\vspace{0.5cm}
\begin{center}
{\Large {\bf Anyonic Interpretation of Virasoro  Characters and the Thermodynamic
Bethe Ansatz} }

\vspace{1.8cm}
{\large A.G. Bytsko$^{*,**}$ ,\ \  A. Fring$^{*}$ }

\vspace{0.5cm}
{\em $*$ Institut f\"ur Theoretische Physik,
Freie Universit\"at Berlin\\ 
Arnimallee 14, D-14195 Berlin, Germany\\ [2mm]
$**$ Steklov Mathematical Institute\\
Fontanka 27, St.Petersburg, 191011, Russia }
\end{center}
\vspace{1.2cm}
 
\renewcommand{\thefootnote}{\arabic{footnote}}
\setcounter{footnote}{0}

\begin{abstract}

Employing factorized versions of characters as products 
of quantum dilogarithms corresponding to irreducible representations
of the Virasoro algebra, we obtain character formulae which admit an anyonic 
quasi-particle interpretation in the context of minimal models in conformal 
field 
theories. We propose anyonic thermodynamic Bethe ansatz equations, together 
with their corresponding equation for the Virasoro central charge, on the base
of an analysis of the  classical limit for the characters and the 
requirement that the scattering matrices are asymptotically phaseless.

\bigskip

\end{abstract}
\vspace{.3cm}
\centerline{November 1997}
\vfill{ \hspace*{-9mm}
\begin{tabular}{l}
\rule{6 cm}{0.05 mm}\\
Bytsko@physik.fu-berlin.de ,\ \ Bytsko@pdmi.ras.ru \\
Fring@physik.fu-berlin.de
\end{tabular}}
\end{titlepage}
\newpage

\section{Introduction}

It is well known, that in 1+1 dimensions spin as well as statistics are
entirely matters of convention. The first observation concerning the spin is
due to Wigner \cite{Wigner}, who found that one particle states always
acquire a form such that they transform as scalars with respect to the
action of the Lorentz group. The second observation concerning the spin is
due to Swieca et al \cite{Swieca}, who provided an explicit form for a
transformation which relates particles obeying different types of
statistics. In general it is possible to make these correspondences
consistent for the bosonic or fermionic situation, whereas for more exotic,
i.e. anyonic, statistics this task is far more complicated.

These statements hold for conformal field theories as well as for massive
models. A natural question which arises, is how these features reflect
itself in the analysis, which relates conformal and massive models, i.e. in
the thermodynamic Bethe ansatz (TBA) \cite{Yang,TBAZam,TBAKM}. Unexpectedly,
it will turn out that, contrary to the fermionic and bosonic situations, in
the anyonic context matters simplify drastically. Clearly, in order to derive
the TBA-equations one should carry out an analysis analogously to the
seminal work by Yang and Yang \cite{Yang}. However, to obtain a first
insight it is also possible to take a shortcut and follow a different route
which reaches the aim quicker.

Recently the observation \cite{KM,KKMM} was made, that the conventional
character formulae for irreducible representations of the Virasoro algebra
of Feign-Fuchs and Rocha-Caridi \cite{FeiginFuchs,Rocha} can be expressed
also in an alternative form. From a mathematical point of view, these
equivalent formulations correspond to generalizations of the famous
identities of Rogers-Ramanujan and Schur \cite{RSR}. From a physical point
of view they are remarkable, since they lead to an interpretation in terms
of fermionic rather than bosonic quasi-particles. Furthermore, they may be
exploited in order to obtain the TBA-equations from their classical limit 
\cite{Rich,KM,Nahm,KKMM}. It is this connection on which we will base our
investigation. More precisely, for the proposed anyonic versions of the
characters, which are equivalent to the fermionic as well as the bosonic
formulae, we assume this connection also to be valid. This assumption is
sustained by our analysis.

Our approach exploits the fact, that, as first observed by Rocha-Caridi \cite
{Rocha}, and further developed by Christe \cite{christe}, certain characters
factorize in terms of infinite products (see also \cite{itz,Cap,Taormina}).
Technically, it turns out to be extremely useful to exploit the properties
of the so-called quantum dilogarithm \cite{Faddeev,Kirilov} in order to deal
with such expressions. In particular the computation of the classical limit
becomes extremely simple in this language.

Our manuscript is organized as follows: In section 2 we make some general
remarks concerning the scattering matrices and the TBA-equations in the
context of generalized, i.e. anyonic, statistics. In section 3 we prove that
certain characters factorize as products of quantum dilogarithms and
investigate their classical limit. In section 4 we propose the generalized
thermodynamic Bethe ansatz equation on the base of a saddle point analysis
and compare with the predictions of section 3. In section 5 we propose some
anyonic type character formulae and discuss  concrete examples of them
together with the explicit construction of some  anyonic quasi-particle
states. We state our conclusions and further outlook in section 6.

\section{On generalized statistics}

In order to motivate our considerations we recall the transformation
proposed in \cite{Swieca}. Assuming, for instance, that $b^{\dagger }(\theta
)$ $\footnote{%
As usual in this context, the two-momentum of a particle is parameterized by
its rapidity $p=m\left( \cosh \theta ,\sinh \theta \right) $.}$ is a
creation operator obeying the equal-time commutation relations for a free
boson one may easily show that 
\begin{equation}
a^{\dagger }(\theta )=b^{\dagger }(\theta )\exp \left( 2\pi
is\int\limits_{\theta }^{\infty }d\theta ^{\prime }b^{\dagger }(\theta
^{\prime })b(\theta ^{\prime })\right)  \label{ant}
\end{equation}
obeys exchange relations related to more exotic (i.e. anyonic) statistics
whenever $s$ is taken not to be an integer or half integer 
\begin{equation}
a^{\dagger }(\theta )a^{\dagger }(\theta ^{\prime })=e^{2\pi is\epsilon
(\theta -\theta ^{\prime })}a^{\dagger }(\theta ^{\prime })a^{\dagger
}(\theta )\quad .  \label{braid}
\end{equation}
Here we have $\epsilon (\theta )=\Theta (\theta )-\Theta
(-\theta )$, where $\Theta (\theta )$ is the usual step-function. Similar
relations also hold for annihilation operators upon replacing $b^{\dagger
}(\theta )$ with $b(\theta )$ and $a^{\dagger }(\theta )$ with $a(\theta )$.
For anyonic statistics these transformations have to be taken with caution,
since as it was already pointed out in \cite{Swieca}, they will lead
immediately to several problems concerning locality, probability
measurements etc. However, these issues are not of our concern here, since
for our purposes we do not need an explicit representation for these
operators. For an introductory account we refer the reader for instance to
some lectures by Montonen \cite{Montonen}.

It is interesting to note, that transformations of the type (\ref{ant})
generalize further and serve as an easy tool to construct explicit
realizations for the Zamolodchikov algebra \cite{AF}, which is generated by
some asymptotic creation operators $Z^{\dagger }(\theta )$.

What the scattering matrices concerns, these transformations may be
incorporated as follows. The factorized n-particle scattering matrix is
defined via the relation 
\begin{equation}
Z_{n}^{\dagger }(\theta _{n})\ldots Z_{1}^{\dagger }(\theta _{1})\left|
0\right\rangle _{\text{out}}=\prod\limits_{1\leq i<j\leq n}S_{ij}(\theta
_{i}-\theta _{j})Z_{1}^{\dagger }(\theta _{1})\ldots Z_{n}^{\dagger }(\theta
_{n})\left| 0\right\rangle _{\text{in}}\,,  \label{defS}
\end{equation}
with $\func{Re}(\theta _{1})>\ldots >\func{Re}(\theta _{n})$. $\left|
0\right\rangle _{\text{in/out}}$ is the vacuum. It follows immediately from
this, that if a state involves two particles which possess the same rapidity
(which is strictly to be understood as a limit $\theta _{i}\rightarrow
\theta _{j}$), $S_{ij}(0)$ will account only for the statistics of the
particles. Sometimes a statistics factor, say $\sigma $, is explicitly
extracted from the S-matrix such that $S_{ij}(0)=1$ always holds. We shall
adopt the convention that this factor is incorporated into $S_{ij}(\theta )$%
. Regarding the scattering matrix in (\ref{defS}) as the one which results
from the bootstrap analysis, it was suggested in \cite{KT,SM} to introduce
instead a matrix 
\begin{equation}
\hat{S}_{ij}(\theta )=S_{ij}(\theta )\left( \exp (-2\pi i\Delta
_{ij}^{+})\Theta (\theta )+\exp (-2\pi i\Delta _{ij}^{-})\Theta (-\theta
)\right) \,,  \label{anS}
\end{equation}
where $\Delta _{ij}^{\pm }$ is the asymptotic phase of $S_{ij}(\theta )$%
\begin{equation}
\lim_{\theta \rightarrow \pm \infty }S_{ij}(\theta )=\exp \left( 2\pi
i\Delta _{ij}^{\pm }\right) \,.
\end{equation}
The expression (\ref{anS}) has the virtue, that it compensates the
asymptotic phases and creates a non-trivial phase at $\theta \rightarrow 0$,
i.e. the anyonic situation. Thus, one observes, that for the scattering
matrix the different statistics may be implemented more easily than for the
asymptotic creation and annihilation operators. It was argued \cite{KT,SM},
that from a physical point of view S-matrices which possess a non-trivial
asymptotic phase should be regarded rather as auxiliary objects. However,
these type of S matrices usually emerge from a bootstrap analysis (see
below). On the other hand, scattering matrices of the type (\ref{anS}) should
be considered as the genuine physical quantities, since they lead to the
correct physical properties, in particular the statistics. From a purely
technical point of view they are however less suitable because of their
non-trivial analytic properties.

Most of the known scattering matrices related to integrable models in 1+1
dimensions exhibit the feature that these asymptotic phases are non-trivial.
Typical examples for diagonal S-matrices, to which we shall be referring
more below, are for instance the ones of affine Toda field theories with
real coupling constants related to simply laced Lie algebras \cite{TodaS} 
\begin{equation}
S_{ij}(\theta )\;=\;\prod_{p=1}^{h}\left\{ 2p-\frac{c(i)+c(j)}{2}\right\}
_{\theta }^{-\frac{1}{2}\lambda _{i}\cdot \sigma ^{p}\gamma _{j}}\;\;.
\label{TodaS}
\end{equation}
The $\{\}_{\theta }$ are building blocks consisting out of sinh-functions,
i.e. $\{x\}_{\theta }=[x]_{\theta }/[x]_{-\theta }$, $[x]_{\theta
}=(\left\langle x+1\right\rangle _{\theta }\left\langle x-1\right\rangle
_{\theta })/(\left\langle x+1-B\right\rangle _{\theta }\left\langle
x-1+B\right\rangle _{\theta })$ and $\left\langle x\right\rangle _{\theta
}=\sinh \frac{1}{2}\left( \theta +\frac{i\pi x}{h}\right) $. $B(\beta )=%
\frac{2\beta ^{2}}{4\pi +\beta ^{2}}$ is the effective coupling constant
which takes values between 0 and 2 when $\beta $ is taken to be purely real. 
$h$ denotes the Coxeter number and $\sigma $ a particular Coxeter element, $%
c(i)=\pm 1$ the colour values related to the bicolouration of the Dynkin
diagram, $\lambda _{i}$ a fundamental weight and $\gamma _{i}$ is $c(i)$
times a simple root. Whenever the coupling constant dependent part is
omitted the S-matrix is referred to as minimal.

The scattering matrix (\ref{TodaS}) has the property $S_{ij}(0)=(-1)^{\delta
_{ij}}$.$\;$ For the phases one obtains in these cases 
\begin{equation}
\Delta _{ij}^{\pm }=\pm \left( \frac{\delta _{ij}}{2}-\left( C^{-1}\right)
_{ij}\right)  \label{minphase}
\end{equation}
for the minimal models ($C_{\text{ }}$denoting here the Cartan matrix of the
corresponding Lie algebra) and 
\begin{equation}
\Delta _{ij}^{\pm }=\mp \frac{\delta _{ij}}{2}
\end{equation}
for the full theory \cite{TBAKM,Dorey}. The Sine-Gordon model, as the first
non-trivial example for a non-diagonal S-matrix also exhibits these
features. The S-matrix reads \cite{ZZ} 
\begin{equation}
S_{SG}(\theta )=\frac{S_{0}(\theta )}{xp-p^{-1}x^{-1}}\left( 
\begin{array}{cccc}
xp-p^{-1}x^{-1} & 0 & 0 & 0 \\ 
0 & p-p^{-1} & x-x^{-1} & 0 \\ 
0 & x-x^{-1} & p-p^{-1} & 0 \\ 
0 & 0 & 0 & xp-p^{-1}x^{-1}
\end{array}
\right) \,
\end{equation}
where $x=\exp (8\pi \theta /\gamma )$, $p=\exp (-i8\pi ^{2}/\gamma )$ and 
\begin{equation}
S_{0}(\theta )=\exp \left( i\int\limits_{0}^{\infty }\frac{dt}{t}\frac{\sin 
\frac{t\theta }{\pi }\sinh t(\frac{\gamma }{16\pi }-\frac{1}{2})}{\sinh 
\frac{t\gamma }{16\pi }\cosh \frac{t}{2}}\right) \,.
\end{equation}
Thus we obtain from this (see e.g. \cite{SM}) 
\begin{equation}
\lim_{\theta \rightarrow \pm \infty }S_{SG}(\theta )=\pm ip^{\pm \frac{1}{2}%
}\left( 
\begin{array}{cccc}
1 & 0 & 0 & 0 \\ 
0 & 0 & p^{\mp 1} & 0 \\ 
0 & p^{\mp 1} & 0 & 0 \\ 
0 & 0 & 0 & 1
\end{array}
\right) \,.
\end{equation}
A testing ground for S-matrices in general is the thermodynamic Bethe ansatz
analysis. This analysis provides information about the consistency of an
S-matrix and leads ultimately to the information to which model a massive
theory flows in the conformal limit. Or vice versa, regarding the massive
model as a perturbation of some conformal field theory in the spirit of \cite
{Zamo1}, it provides the information about the conformal ancestor of a given
massive theory. The key equations of the thermodynamic Bethe ansatz \cite
{Yang,TBAZam,TBAKM} related to diagonal S-matrices, which determine the
pseudo-energies $\varepsilon _{i}$ are 
\begin{equation}
\varepsilon _{i}=\pm \sum\limits_{j=1}^{r}N_{ij}\ln (1\pm e^{-\varepsilon
_{j}})\quad ,  \label{TBA1}
\end{equation}
where the upper and lower signs correspond to the fermionic and bosonic
versions, respectively. $r$ is the number of particles and the matrix $%
N_{ij} $ is related to the phases introduced above 
\begin{equation}
N_{ij}=\Delta _{ij}^{-}-\Delta _{ij}^{+}\quad .  \label{Nmat}
\end{equation}
The central charge of the Virasoro algebra of the underlying conformal field
theory is given in the following form

\begin{equation}
c=\frac{6}{\pi ^{2}}\sum\limits_{i=1}^{r}\left\{ 
\begin{array}{l}
L\left( \frac{1}{1+e^{\varepsilon _{i}}}\right) \qquad \,\text{fermionic} \\ 
L\left( e^{-\varepsilon _{i}}\right) \qquad \quad \,\text{bosonic}
\end{array}
\qquad \right. \quad .  \label{TBA2}
\end{equation}
Here $L(x)$ $=\sum\limits_{n=1}^{\infty }\frac{x^{n}}{n^{2}}+\frac{1}{2}\ln
x\ln (1-x)$ denotes the Rogers dilogarithm \cite{Lewin}.

Recently the remarkable observation was made, that for particular models the
fermionic TBA-equations also result from a limiting procedure when
considering characters of the Virasoro algebra related to certain conformal
field theories. It will be our goal to extent this analysis to generalized,
i.e. anyonic, statistics.

\section{Factorized Characters and Quantum Dilogarithms}

The bosonic realization \cite{FeiginFuchs,Rocha} of the character of an
irreducible representation of the Virasoro algebra related to the Hilbert
space of the minimal models ${\cal M}(s,t)$ \cite{BPZ} with central charge $%
c=1-\frac{6(s-t)^{2}}{s\;t}$ and highest weight $h_{n,m}=\frac{%
(nt-ms)^{2}-(s-t)^{2}}{4\;s\;t}$ reads 
\begin{eqnarray}
\chi _{n,m}^{s,t}(q) &=&\frac{q^{h_{n,m}-\frac{c}{24}}}{(q)_{\infty }}%
\sum_{k=-\infty }^{\infty }q^{stk^{2}}\left(
q^{k(nt-ms)}-q^{k(nt+ms)+nm}\right) \;\;  \label{Bose} \\
&=&\frac{q^{h_{n,m}-\frac{c}{24}}}{(q)_{\infty }}\hat{\chi}%
_{n,m}^{s,t}(q)=q^{h_{n,m}-\frac{c}{24}}\sum\limits_{k=0}^{\infty
}c_{k}q^{k}\,.  \label{Bose2}
\end{eqnarray}
Irreducibility also demands that $s$ and $t$ are co-prime and $1\leq n\leq
s-1 $ and $1\leq m\leq t-1$. We have used the standard abbreviation for
Euler's function $(q)_{m}=\prod\limits_{k=1}^{m}(1-q^{k})$ and $(q)_{0}=1$. 
In the following
analysis we shall be exploiting the relation of the expressions (\ref{Bose}%
) to the quantum dilogarithm whose defining relations are 
\begin{equation}
\ln _{q}(\theta ):=\prod\limits_{k=0}^{\infty }(1-e^{2\pi i\theta
}q^{k})\,=\exp \sum\limits_{k=1}^{\infty }\frac{1}{k}\frac{e^{2\pi i\theta k}%
}{q^{k}-1}\,.  \label{quprod}
\end{equation}
Taking $q=e^{2\pi i\tau },$ we assume that $\func{Im}(\tau )>0$ in order to
guarantee the convergence of (\ref{quprod}). We see from (\ref{quprod}) that 
$\ln _{q}(\theta )$ is a pseudo-double-periodic function 
\begin{equation}
\ln _{q}(\theta +1)=\ln _{q}(\theta )\qquad \hbox{and}\qquad \ln _{q}\left(
\theta +\tau \right) =\frac{1}{1-e^{2\pi i\theta }}\ln _{q}(\theta )\quad .
\label{qf1}
\end{equation}
It follows easily from this that 
\begin{equation}
\ln _{q}(\theta )=\sum\limits_{k=0}^{\infty }\frac{(-1)^{k}q^{\frac{%
k\left( k-1\right) }{2}}e^{2\pi i\theta k}}{\left( q\right) _{k}}\quad
\label{notinv}
\end{equation}
and 
\begin{equation}
\frac{1}{\ln _{q}(\theta )}=\sum\limits_{k=0}^{\infty }\frac{e^{2\pi
i\theta k}}{\left( q\right) _{k}}\quad .  \label{inv}
\end{equation}
The function $\ln _{q}(\theta )$ was coined as quantum dilogarithm because
in its classical limit $\tau \rightarrow 0$ ($\tau $ may be thought of as $%
\hbar $ or $1/L$, with $L$ being the size of the system (see section 5)) 
the singular term involves Euler's dilogarithm $Li_{2}\left(
x\right) =\sum\limits_{n=1}^{\infty }\frac{x^{n}}{n^{2}}$ \cite{Lewin} 
\begin{equation}
\lim\limits_{\tau \rightarrow 0}\ln _{q}(\theta )\;=\;\exp \left\{ \frac{1}{%
2\pi i\tau }\;Li_{2}\left( e^{2\pi i\theta }\right) \;+\;{\cal {O}}\left(
1\right) \right\} \;\;\;.  \label{dilimit}
\end{equation}
This follows directly from (\ref{quprod}). A relation between characters and
the quantum dilogarithm in the form of a sum as (\ref{Bose}) is most easily
established by re-expressing them in terms of theta functions, which in turn
are composed of three quantum dilogarithms. However, this type of relation
is of no concern to us here, instead we will seek to factorize the quantity $%
\hat{\chi}_{n,m}^{s,t}(q)$ in the following form

\begin{equation}
\hat{\chi}_{n,m}^{s,t}(q)=\prod\limits_{k=1}^{N}\ln _{q^{b}}(a_{k})\quad
\quad .\quad  \label{dilfac}
\end{equation}
The advantage of this realization is twofold. On one hand it will serve us
to obtain anyonic versions of the characters and on the other it may be used
to carry out the classical limit effortless. Introducing the S-modular
transformed version of $q,$ i.e. $\hat{q}=\exp (-2\pi i/\tau ),$ we carry
out the classical limit $q\rightarrow 1^{-}$ (and hence $\hat{q}\rightarrow
0^{+}$) by means of (\ref{dilimit}) 
\begin{equation}
\lim_{\tau \rightarrow 0}\chi _{n,m}^{s,t}(q)=\hat{q}^{\frac{Li_{2}(1)}{4\pi
^{2}}\left( \frac{N}{b}-1\right) }=\hat{q}^{\frac{1}{24}\left( \frac{N}{b}%
-1\right) }\quad .  \label{limcc}
\end{equation}
The -1 in (\ref{limcc}) occurs due to the fact that $(q)_{\infty }$, which
appears in (\ref{Bose}), coincides with $\ln _{q}(\tau )$ and therefore its
limit is also ruled by (\ref{dilimit}).

It is clear, that in general the classical limit for expressions like (\ref
{Bose}) is quite non-trivial to carry out directly. However, one may exploit
the behaviour of these formulae under the modular transformation. It is well
known, that the characters transform under the S-modular transformation in
the following general form \cite{itz} 
\begin{equation}
\chi _{n,m}^{s,t}(q)=\sum\limits_{n^{\prime },m^{\prime }}C_{nm}^{n^{\prime
}m^{\prime }}\chi _{n^{\prime },m^{\prime }}^{s,t}(\hat{q})\quad .
\end{equation}
The $C_{nm}^{n^{\prime }m^{\prime }}$ are explicitly known constants. In the
classical limit we obtain 
\begin{eqnarray}
\lim_{\tau \rightarrow 0}\chi _{n,m}^{s,t}(q) &=&\sum\limits_{n^{\prime
},m^{\prime }}\hat{q}^{h_{n^{\prime },m^{\prime }}-\frac{c}{24}%
}\,C_{nm}^{n^{\prime }m^{\prime }} \\
&=&C_{nm}^{\bar{n}\bar{m}}\hat{q}^{-\frac{c_{eff}}{24}}\left(
1+\sum\limits_{n^{\prime },m^{\prime }\neq \bar{n},\bar{m}}\hat{q}%
^{h_{n^{\prime },m^{\prime }}-h_{\bar{n},\bar{m}}}\,C_{nm}^{n^{\prime
}m^{\prime }}/C_{nm}^{\bar{n}\bar{m}}\,\right) .  \label{modlim}
\end{eqnarray}
Here we have introduced the so-called effective central charge 
\begin{equation}
c_{eff}=c-24h_{\bar{n},\bar{m}}=1-\frac{6(\bar{n}t-\bar{m}s)^{2}}{\;s\;t}%
\quad ,  \label{eff}
\end{equation}
where $h_{\bar{n},\bar{m}}$ denotes the lowest of all conformal weights in
the model. Comparison with (\ref{limcc}) yields the following constraint on
the number of factors $N$ and the constant $b$%
\begin{equation}
c_{eff}=1-\frac{N}{b}\, \; , \label{eeff}
\end{equation}
for every character which factorizes in the form of (\ref{dilfac}). These
formulae provide a criterion which serves to decide whether any given
character of the general Feigin-Fuchs/Rocha-Caridi form (\ref{Bose}) may be
factorized. From a mathematical point of view it provides an existence
criterion for certain Rogers-Ramanujan-Schur type identities \cite
{RSR,Hardy,Niven}.

We shall now provide explicit examples for factorized characters of the form
(\ref{dilfac}). We commence by considering the well-known Gau\ss -Jacobi
identity (see for instance \cite{Kirilov}) 
\begin{equation}
\sum_{k=-\infty }^{\infty }(-1)^{k}x^{\frac{k(k+1)}{2}}y^{\frac{k(k-1)}{2}%
}=\prod\limits_{k=1}^{\infty }(1-x^{k}y^{k})(1-x^{k-1}y^{k})(1-x^{k}y^{k-1}).
\end{equation}
We attempt to match the left hand side of this formula with $\hat{\chi}%
_{n,m}^{s,t}(q)$ in the form of (\ref{Bose}). Making the ansatz $x=q^{a}$
and $y=q^{b}$ one easily derives that the only solution\footnote{%
More precisely, one also obtains the alternative solution with the condition 
$t=2m$. However, from a physical point of view these solutions are
equivalent because of the symmetry $\chi _{n,m}^{s,t}(q)=\chi
_{m,n}^{t,s}(q)$.} is $a=nm$ and $b=\frac{st}{2}-nm$ together with the
condition $s=2n$. We therefore obtain 
\begin{equation}
\hat{\chi}_{n,m}^{2n,t}(q)=\prod\limits_{l\in {\bf e}^{2n}}\ln
_{q^{nt}}(\tau l)\quad ,  \label{ch1}
\end{equation}
where we introduced the set {\bf e}$^{2n}=\{nm,nt-nm,tn\}$. This is a
well-known formula and may be found for instance in \cite{itz,christe}. We
may also carry out a consistency check by taking the limit $\lim_{\tau
\rightarrow 0}\hat{\chi}_{n,m}^{2n,t}(q)$ and compare (\ref{eff}) with (\ref
{eeff}). We obtain for the effective central charge 
\begin{equation}
c_{eff}=1-\frac{3}{nt}=1-\frac{3(\bar{n}t-2\bar{m}n)^{2}}{nt}\quad .
\label{ccc}
\end{equation}
This is only possible if the quantity $\lambda =(\bar{n}t-2\bar{m}n)^{2}=1$,
which is indeed the case as a simple argument shows. The lowest value $%
\lambda $ can take is zero, which is obviously impossible, since by
construction $s$ and $t$ are to be co-prime. A theorem concerning the
Diophantine equations in number theory (e.g. Theorem 1.3 in \cite{Niven})
states, that the equation $t\bar{n}-s\bar{m}= \pm 1$ possess always an integer
solution when $t$ and $s$ are co-prime. One may also show that (e.g. \cite
{Niven} p.96) $0\leq \bar{n}<s$ and $0\leq \bar{m}<t.$ Therefore we have $%
\lambda =1,$ together with the restrictions on $\bar{n}$ and $\bar{m}$.
Hence, equation (\ref{ccc}) holds.\footnote{%
Reversing here the argumentation, we have also proven a theorem from number
theory by a physical reasoning as a by-product of our analysis of 
characters.}

According to the arguments of Christe \cite{christe} there is only one other
class of characters which also factorizes in the form of (\ref{dilfac}),
that is $\hat{\chi}_{n,m}^{3n,t}(q)$. To our knowledge this was only
established via computer
\footnote{After submission of this manuscript, it was brought to our
attention by W. Eholzer that there exist analytical 
arguments \cite{KRV,ES}. Our analysis is close to \cite{ES}.}
on a case-by-case level \cite{christe}. We prove
this assertion by considering Watson's identity \cite{Watson}

\begin{eqnarray}
\sum_{k=-\infty }^{\infty }x^{\frac{3k^{2}+k}{2}%
}y^{3k^{2}}(y^{-2k}-y^{4k+1}) &=&\prod\limits_{k=1}^{\infty
}(1-x^{k}y^{2k})(1-x^{k}y^{2k-1})(1-x^{k-1}y^{2k-1})  \nonumber \\
&&\times (1-x^{2k-1}y^{4k-4})(1-x^{2k-1}y^{4k})\quad .
\end{eqnarray}

Making once more an ansatz of the type $x=q^{a}$ and $y=q^{b}$ and trying to
match with $\hat{\chi}_{n,m}^{s,t}(q),$ we obtain now that the only solution%
\footnote{%
Once more, one also obtains alternative solutions, which are equivalent
because of the symmetries $\chi _{n,m}^{s,t}(q)=\chi _{m,n}^{t,s}(q)=\chi
_{s-n,t-m}^{s,t}(q)$.} is $a=\frac{2}{3}st-2nm$ and $b=nm$ together with the
condition $s=3n$, such that 
\begin{equation}
\hat{\chi}_{n,m}^{3n,t}(q)=\prod\limits_{l\in {\bf e}^{3n}}\ln
_{q^{4nt}}(\tau l)\quad .\quad  \label{ch2}
\end{equation}
Here we employ the set {\bf e}$^{3n}=%
\{nm,2nt,2n(t-m),n(2t-m),n(2t+m),2n(t+m),n(4t-m),4nt)\}$.

In this case we obtain from the limit $\lim_{\tau \rightarrow 0}\hat{\chi}%
_{n,m}^{3n,t}(q)$ for the effective central charge 
\begin{equation}
c_{eff}=1-\frac{8}{4nt}=1-\frac{2(\bar{n}t-3\bar{m}n)^{2}}{nt}\quad ,
\end{equation}
which is only satisfied if we have $\lambda ^{\prime }= (\bar{n}t-3\bar{m}%
n)^2 = 1$. By the same reasoning as before we deduce also in this case that
indeed $\lambda ^{\prime }=1$.

We would like to stress, that despite the fact that characters of the type (%
\ref{ch1}) and (\ref{ch2}) may appear somewhat exotic, they are ubiquitous.
All unitary minimal models, i.e. ${\cal M}(n,n+1)$, possess factorizable
sectors. Moreover, for minimal models of the type ${\cal M}(2,t)$ and ${\cal %
M}(3,t)$ \underline{all} sectors are of this form.

We conclude this section with the remark, that the argument of Christe
concerning the non-existence of other factorizable characters is based on
the assumption that the characters factorize precisely in the form (\ref
{dilfac}). However, with some minor modifications our analysis works equally
well for characters of the form $\hat{\chi}_{n,m}^{s,t}(q)=\prod%
\limits_{k=1}^{N}\ln _{q^{b}}(a_{k})/\prod\limits_{l=1}^{N^{\prime }}\ln
_{q^{b^{\prime }}}(a_{l}^{\prime }).$ At first sight, such expressions occur
in the work of \cite{Rocha,itz,Cap,Taormina}, but it turns out that they are
always ``reducible'' to (\ref{dilfac}) (see below). It is left for future
investigations to extend our considerations to characters which factorize
genuinely different from (\ref{dilfac}).

\section{From characters to TBA}

In the preceding section we have demonstrated that the classical limit $%
q\rightarrow 1^{-}$ may be carried out effortless once a character
factorizes. On the other hand, there exists a method \cite{Rich,KM,Nahm,KKMM}
which allows to perform the limiting procedure for every type of character.

Before investigating characters in particular, let us consider the function 
\begin{equation}
\chi (q)=\sum_{\vec{m}=0}^{\infty }\frac{q^{\vec{m}A\vec{m}^{t}+\vec{m}\cdot 
\vec{B}}}{(q^{b})_{m_{1}}\ldots (q^{b})_{m_{r}}}\quad .  \label{genchar}
\end{equation}
$A$ is assumed to be an ($r\times r)$-matrix and $\vec{B}$ is an $r$%
-component vector. Despite the suggestive notation, $\chi (q)$ does not need
to be a character for the purpose of deriving an equations of TBA type.
However, for particular $A$ and $\vec{B}$ equations (\ref{genchar}) coincide
(up to the factor $q^{h_{n,m}-\frac{c}{24}}$) with certain 
Virasoro characters, which give rise to a remarkable fermionic
quasi-particle interpretation \cite{KM,KKMM,BF}. Below we will demonstrate,
that whenever $b$ is taken to be an integer greater than one and $\chi (q)$
in (\ref{genchar}) coincides with a generating function for characters, it
admits an anyonic interpretation. We will now apply the method of
 \cite{Rich,KM,Nahm,KKMM}  to an expression of
the type (\ref{genchar}). Hitherto, this has been carried out for the
particular situation when $b=1,A\neq 0$ and $\vec{B}=0$. When $\vec{B}$ only
characterizes a different superselection sector it is justified to set it to
zero. We will include it into our considerations. Furthermore, since we are
interested in anyonic representations, we only demand $b$ to be a positive
integer.

\subsection{Saddle Point Analysis}

Viewing the series in (\ref{Bose2}) as a Laurent expansion and assuming that 
$A$ is a positive definite matrix, a comparison between (\ref{Bose2}) and (%
\ref{genchar}) together with the application of Cauchy's theorem yields 
\begin{equation}
c_{k-1}=\sum\limits_{\vec{m}}\oint \frac{dz}{2\pi i}\frac{z^{\vec{m}A\vec{m}%
^{t}+\vec{B}\cdot \vec{m}-k}}{\left( z^{b}\right) _{m_{1}}\ldots \left(
z^{b}\right) _{m_{r}}}\simeq \int d\vec{m}\oint \frac{dz}{2\pi i}\frac{z^{%
\vec{m}A\vec{m}^{t}+\vec{B}\cdot \vec{m}-k}}{\left( z^{b}\right)
_{m_{1}}\ldots \left( z^{b}\right) _{m_{r}}}.  \label{ck}
\end{equation}
As a first approximation we have changed the $r$ sums over $m_{i}$ into
integrals. This step also makes possible restrictions, which typically occur
for fermionic type representations (see section 5), insignificant. A natural
setting which gives information about the asymptotic behaviour is the saddle
point approximation or method of the steepest descent, which is well known
from the theory of complex functions. In order to be able to apply the
saddle point method to (\ref{ck}) we need the integrand to acquire the form
of an exponential 
\begin{equation}
c_{k-1}\simeq \int d\vec{m}\oint \frac{dz}{2\pi i}\exp 
 \left[ f(\vec{m},z) \right] ,
\end{equation}
with 
\begin{equation} \label{fm}
 f(\vec{m},z)=\left( \vec{m}A\vec{m}^{t}+\vec{B}\cdot 
 \vec{m}-k\right) \ln z -
 \sum\limits_{l=1}^{r}\int\limits_{0}^{m_{l}}dt\ln \left( 1-z^{bt}\right) .
\end{equation}
Here we have approximated once more a sum by an integral, i.e. $\ln
(z)_{m}=\sum\limits_{k=1}^{m}\ln \left( 1-z^{k}\right) \simeq
\int\limits_{0}^{m}dt\ln \left( 1-z^{t}\right) $. The saddle point
conditions with respect to the integration in $m_{i}$ and $z$ are 
\begin{equation}
\partial _{m_{i}}f(\vec{m},z)|_{m_{i}=n_{i}}=0\qquad \text{and\qquad }%
\partial _{z}f(\vec{m},z)|_{z_{0}}=0\,.  \label{saddle}
\end{equation}
So, $f(\vec{n},z_{0})$ is the value of the function at the saddle point $(%
\vec{n},z_{0})$. The relations resulting from the first constraining
equations read 
\begin{equation}
z^{B_{i}+\sum_{j}\left( A_{ij}+A_{ji}\right) n_{j}}+z^{bn_{i}}=1\qquad 
\text{for}\quad i=1,\ldots ,r\,.  \label{gTBA}
\end{equation}
These equations serve to fix $\vec{n}$. We obtain from (\ref{gTBA}) 
\begin{equation} \label{nan}
 \left( 2\vec{n}A\vec{n}^{t}+\vec{B}\cdot \vec{n}\right)\,(\ln z)^2 =
 \sum\limits_{l=1}^{r} (\ln z^{n_{l}}) \ln (1-z^{bn_{l}})\,.
\end{equation}
Substituting this relation into (\ref{fm}) and exploiting the
following identity \cite{Lewin} for Roger's dilogarithm 
\begin{equation}
 (\ln z) \, \int\limits_{0}^{m}dt\ln \left( 1-z^{t}\right) =
 L(1-z^{m})+\frac{1}{2} (\ln z^{m}) \ln (1-z^{m}) \, , 
\end{equation}
we derive 
\begin{equation}
 f(\vec{n},z)=-\left( k-\frac{1}{2}\vec{B} \cdot \vec{n}\right) \ln z-
 \frac{1}{b\ln z}\sum\limits_{i=1}^{r}L(1-z^{bn_{i}})\,.
\end{equation}
Using the formula $\frac{d}{dz}L(z)=-\frac{1}{2}\left( \frac{\ln (1-z)}{z}+%
\frac{\ln z}{1-z}\right) $, we obtain the remaining saddle point condition
in (\ref{saddle}) 
\[
 k\left( \ln z_{0}\right) ^{2}=\frac{1}{b}\sum\limits_{i=1}^{r}
 L(1-z_{0}^{bn_{i}})-\left( \ln z_{0}\right) ^{2}\left( 
 \vec{n}A\vec{n}^{t}+\frac{b}{2}\sum_{i=1}^{r}n_{i}^{2}z_{0}^{bn_{i}-B_{i}-
 \sum_j (A_{ij}+A_{ji})n_{j}}\right) \, . 
\]
Analyzing the limit $k\rightarrow \infty$ of this relation
with the help of (\ref{gTBA})--(\ref{nan}), we infer that
\begin{equation}
\ln z_{0}=-\sqrt{{\cal K}/k}\left( 1+{\cal %
O}(k^{-1})\right) \quad \quad \text{{\rm with}}\quad {\cal K}=\frac{1}{b}%
\sum\limits_{i=1}^{r}L(1-z_{0}^{bn_{i}})\quad .  \label{kkk}
\end{equation}
That is, $z_{0\text{ }}$tends to one for large $k$.
Thus, the asymptotics of $c_k$ is given by
\begin{equation}
 c_{k}\simeq \exp \left[ f(\vec{n},z_{0})\right] \simeq \exp 
 \left( 2\sqrt{k{\cal K}}\right) \ .  \label{cccc}
\end{equation}
Now we are in a position to perform the classical limit. In \cite
{Rich,KM,Nahm,KKMM} it was observed, that once the coefficients in the
Laurent expansion are of the form (\ref{cccc}) one may approximate 
\begin{equation}
\sum\limits_{k=0}^{\infty }c_{k}q^{k}\simeq \int\limits_{0}^{\infty
}dk\,c_{k}q^{k}\simeq \exp \left( \frac{i{\cal K}}{2\pi \tau }+{\cal O}(\ln
\tau )\right) \,.
\end{equation}
A comparison with (\ref{modlim}) leads to the following formula for the
effective central charge 
\begin{equation}
c_{eff}=\frac{6}{\pi ^{2}}{\cal K\,}\quad .  \label{efff}
\end{equation}

\subsection{The TBA equation}

Introducing now the quantity $\xi _{i}=1-z_{0}^{bn_{i}}$ we obtain from (\ref
{kkk}) and (\ref{efff}) for the central effective charge 
\begin{equation}
c_{eff}=\frac{6}{b\pi ^{2}}\sum\limits_{i=1}^{r}L\left( \xi _{i}\right)
\quad , \label{ceffe}
\end{equation}
where the quantities $\xi _{i}$ are determined by the equations resulting
from (\ref{gTBA}) 
\begin{equation}
\xi _{i}^{b}=z_{0}^{bB_{i}}\prod\limits_{j=1}^{r}(1-\xi
_{j})^{(A_{ij}+A_{ji})}\quad .  \label{TBA}
\end{equation}

As argued in the preceding subsection $z_{0\text{ }}$tends to one and
different choices for the vector $\vec{B}$, i.e. different superselection
sectors, have no effect on the value of the effective central charge. Taking
 $A$ to be of the form 
\begin{equation}
A_{ij}=\frac{N_{ij}}{2}+\left\{ 
\begin{array}{ll}
b \frac{ \delta _{ij}}{2} \qquad& \text{fermionic} \\
0 & \text{bosonic}  
\end{array}
\right. \qquad  \label{Amat}
\end{equation}
and relating $\xi _{i}$ to the pseudo-energies as 
\begin{equation}
\xi _{i}=\left\{ 
\begin{array}{ll}
\frac{1}{1+e^{\varepsilon _{i}}}  \qquad &\text{fermionic} \\
e^{-\varepsilon _{i}} & \text{bosonic} 
\end{array}
\right. \qquad \qquad ,  \label{pseudo}
\end{equation}
one recovers the known fermionic and bosonic TBA-equations (\ref{TBA1})
and (\ref{TBA2}), upon choosing $b=1$.

The analysis for particular fermionic cases was carried out in \cite
{Rich,KM,Nahm,KKMM}. One might be surprised, that formally a fermionic type
expression gives rise to a fermionic as well as a bosonic TBA-equation. Such
features are however common in the TBA analysis itself, in which also
bosonic statistics may correspond to fermionic boundary conditions and
vice versa.

As explained in section 2, anyonic statistics may be implemented into the
scattering theory by removing the asymptotic phases in the S-matrix and
introducing them at $\theta =0$. As a consequence the quantity $N_{ij}$ (\ref
{Nmat}) becomes zero. In this case a substitution of (\ref{Amat}) into (\ref
{TBA}) leads to 
\begin{equation}
\xi _{i}= \frac{1}{2} \qquad \text{and \qquad }\xi _{i}=1 \,  \label{anTBA}
\end{equation}
as two possible versions of anyonic TBA-equations. Notice, that the anyonic
TBA-equations (\ref{anTBA}) correspond precisely to the situation in which 
the pseudo-energies are taken to be zero. 
The contribution to 
the effective central charge from each particle is in these cases either
$\frac{1}{b}$ or $\frac{1}{2b}$, according to  (\ref{ceffe}) (recall that
$L(1)= 2 L(1/2)= \pi^2/6$). So, in some sense the anyons carry a remembrance
on a fermionic or bosonic nature. Below we will see that this picture may
also be related to  Pauli's exclusion principle. 

In both cases the complicated
coupling between equations involving different types of particles has
vanished. This is a  virtue of the anyonic TBA-equation, in which all
particles of the same kind contribute 
equally to the central charge.

To summarize: we have proposed a transformation either from a bosonic or
fermionic type of TBA-equation to an anyonic one.  Of course this 
transformation has to preserve the value of the central charge. Therefore, 
we expect the existence of character formulae of
the type (\ref{genchar}), admitting two forms, i.e.  involving a matrix 
of the type (\ref{Amat}) either
with $N_{ij}\neq 0$ and $b=1$ or $N_{ij}=0$ and $b\neq 1$.

\section{Anyonic Characters and States}

As already mentioned, when interpreted as partition functions, the character
formulae of the preceding sections possess some fermionic realizations.
These representations emerge in two different types. One of them is
intimately related to simply laced Lie algebras in the following sense. When 
$A$ is chosen to be the inverse of the Cartan matrix and $b=1$, (\ref{genchar})
coincides with some characters corresponding to minimal models of conformal
field theory. The number of fermions contained in the model equals the rank $%
r$ of the Lie algebra in this case. The summation over $m_{1},m_{2},\ldots
,m_{r}$ may be restricted in some way, indicating that certain particles may
only appear in conjunction with others. From a Lie algebraic point of view
this expresses usually some symmetry in the Dynkin diagram. The vector $\vec{%
B}$ characterizes different superselection sectors for a particular theory.

Notice that such formulae, i.e. $A$ being the inverse of the Cartan matrix,
correspond precisely to the fermionic case in (\ref{Amat}), if the matrix $%
N_{ij}$ (\ref{Nmat}) is defined via the asymptotic phases of the minimal
affine Toda field theory (\ref{minphase}). It is very interesting to note
that some models possess different fermionic realizations, for example, the
Ising model with $c=1/2$ can be related either to $E_{8\text{ }}$or to $%
A_{1} $, which indicates the possible relevant perturbations already at the
conformal level.

\subsection{Anyonic Characters}

According to the ideology laid down above, one might expect the existence of
character formulae corresponding to the same models, which admit an anyonic
quasi-particle interpretation. In this section we will demonstrate how to
obtain them and show that these formulae as well relate to simply laced Lie
algebras.

First of all we assume the main part of the characters, $\hat{\chi}$, to
factorize in the form of (\ref{dilfac}). With $a_{l}=\tau l$ and $l$ being
element in the $N$-dimensional set ${\bf e}$, we obtain 
\begin{equation}
\chi _{n,m}^{s,t}(q)=q^{h_{n,m}-\frac{c}{24}}\prod\limits\Sb l=1  \\ l\notin 
{\bf e}  \endSb ^{b}\left( \ln _{q^{b}}(\tau l)\right) ^{-1}.  \label{anprod}
\end{equation}
Employing (\ref{inv}) in (\ref{anprod}), we obtain the desired anyonic
realizations

\begin{equation}
\chi _{n,m}^{s,t}(q)=q^{h_{n,m}-\frac{c}{24}}
\sum\limits_{\vec{k}=0}^{\infty} \frac{q^{%
\vec{k}\cdot \vec{B}_{n,m}^{s,t}}}{\left( q^{b}\right) _{k_{1}}\ldots \left(
q^{b}\right) _{k_{(b-N)}}}.  \label{anychar}
\end{equation}

We have introduced here the ($b-N$)-dimensional vector $\vec{B}%
_{n,m}^{s,t}=\{1,2,3,\ldots ,b\}/${\bf e}$.$ Our particular examples of
section 3 for factorizing characters (\ref{ch1}) and (\ref{ch2}) acquire the
following form 
\begin{eqnarray}
\chi _{n,m}^{2n,t}(q) &=&q^{h_{n,m}-\frac{c}{24}}\sum\limits_{\vec{k}}\frac{%
q^{\vec{k}\cdot \vec{B}_{n,m}^{2n,t}}}{\left( q^{nt}\right) _{k_{1}}\ldots
\left( q^{nt}\right) _{k_{nt-3}}}  \quad ,  \\
\chi _{n,m}^{3n,t}(q) &=&q^{h_{n,m}-\frac{c}{24}}\sum\limits_{\vec{k}}\frac{%
q^{\vec{k}\cdot \vec{B}_{n,m}^{3n,t}}}{\left( q^{4nt}\right) _{k_{1}}\ldots
\left( q^{4nt}\right) _{k_{4nt-8}}}\quad ,
\end{eqnarray}
with $\vec{B}_{n,m}^{2n,t}=\{1,2,3,\ldots ,nt\}/${\bf e}$^{2n}$ and $\vec{B}%
_{n,m}^{3n,t}=\{1,2,3,\ldots ,4nt\}/${\bf e}$^{3n}.$

Apparently (\ref{anprod}) and (\ref{anychar}) correspond to the second
case in (\ref{Amat}), since all $A_{ij}=0$. The first case can be recovered 
if one considers a character of the type
\begin{equation}
\chi(q)=  \frac{\prod\limits_l^N \ln_{q^b}(\tau l + \frac{1}{2})}
{\prod\limits_{l'}^{N'} \ln_{q^b}(\tau l' )} \qquad .  
\label{TYPE2}
\end{equation}
Employing then (\ref{notinv}) and  (\ref{inv}), we can rewrite this 
factorized character in the form 
\begin{equation}
\chi (q)=\sum_{\vec{k}=0}^{\infty }\frac{q^{ \frac{b}{2}
({k_1}^{2}+ \ldots + k_N^2 )    +
\vec{k}\cdot \vec{B}}}{(q^{b})_{k_{1}}\ldots 
(q^{b})_{k_{N+N'}}}  \quad .  
\label{gencharac}
\end{equation}
Below we will encounter an example of this type.

\subsection{Anyonic Quasi-particles}

Anyonic quasi-particle states may be constructed analogously to fermionic
states as proposed by the Stony Brook group \cite{KM,KKMM}. First of all one
regards the characters as partition functions 
\begin{equation}
\chi \left( q=e^{-\frac{2\pi v}{LkT}}\right) \sim Z=\sum_{{\rm states}}e^{-%
\frac{{\rm E(states)}}{kT}}=\sum_{l=0}^{\infty }P(E_{l})e^{-\frac{E_{l}}{kT}%
}\quad .  \label{part}
\end{equation}
$k$ denotes here Boltzmann's constant, T the temperature, L the size of a
box which serves to quantize the momenta, $v$ the speed of sound, $E_{l\text{
}}$the energy of a particular level and $P(E_{l})$ its degeneracy. The aim
is now to bring (\ref{anychar}) into the form of the right hand side of (\ref
{part}). A direct comparison then leads to the construction of
quasi-particle states 
\begin{equation}
|p_{1}^{1},\ldots ,p_{1}^{k_{1}},\ldots ,p_{b-N}^{1},\ldots
,p_{b-N}^{k_{n}}\rangle \;\;,
\end{equation}
which are in one-to-one correspondence to the states of the Virasoro
algebra. The energy of each of these states can then be decomposed further
into the contributions resulting from each single quasi-particle 
\begin{equation}
E_{l}\;=\;\sum_{a=1}^{b-N}\sum_{i_{a}=1}^{k_{a}}\;e_{a}(p_{a}^{i_{a}})\qquad
,\qquad p_{l}\;=\;\sum_{a=1}^{b-N}\sum_{i_{a}=1}^{k_{a}}\;p_{a}^{i_{a}}.
\label{quasp}
\end{equation}
The dispersion relation is to be understood as $e_{a}(p_{a}^{i})=v\left|
p_{a}^{i}\right| $.

Technically, one has several options to replace the factors $\frac{1}{%
(q)_{k_{i}}}$ in order to achieve formal equality between (\ref{anychar})
and (\ref{part}). In the fermionic case one makes use of the generating
function \cite{Hardy,Niven} 
\begin{equation}
\sum_{n=0}^{\infty }{\cal P}(n,m)q^{n}\;=\;\frac{q^{\frac{1}{2}m(m-1)}}{%
(q)_{m}},  \label{party}
\end{equation}
for the number of partitions ${\cal P}(n,m)$ of a positive integer $n$ into $%
m$ \underline{distinct} non-negative integers.\footnote{%
There exists also the possibility to employ the generating function for the
number of partitions of a positive integer $n$ into $m$ distinct
non-negative integers smaller than $M$, which gives rise to fermionic
quasi-particle states of a different nature \cite{KKMM}.} Performing the
same manipulations on the anyonic type character (\ref{anychar}), we obtain
for the anyonic single quasi-particle momenta (in units of $\frac{2\pi }{L}$%
) 
\begin{equation}
p_{a}^{i}=(B_{n,m}^{s,t})_{a}+\frac{b}{2}-\frac{b}{2}%
\sum_{l=1}^{b-N}k_{l}+bN_{a}^{i} .  \label{quasis}
\end{equation}
Here the $N_{a}^{i}$ are distinct positive integers, and therefore we still
have, despite the the anyonic nature of our quasi-particles, a Pauli type
exclusion principle. 

Since here we are considering anyonic quasi-particle states, we can relax
the requirement that the partitions have to be carried out into distinct
numbers. In this case we may also use the generating function 
\begin{equation}
\sum\limits_{n=0}^{\infty }{\cal Q}(n,m)q^{n}=\frac{1}{(q)_{m}}\quad
\label{part2}
\end{equation}
for the number of partitions ${\cal Q}(n,m)$ of a positive integer $n$ into
positive integers smaller than $m$. In this way we obtain an alternative
formula for the single quasi-particle momenta (in units of $\frac{2\pi }{L}$%
) 
\begin{equation}
p_{a}^{i}=(B_{n,m}^{s,t})_{a}+bN_{a}^{i} \;\; ,
\end{equation}
with $N_{a}^{i}$ being some positive integers.

From a formal mathematical point of view we do not know which choice of the
partition functions (\ref{party}) or (\ref{part2}) is adequate. However, it
seems plausible, that in agreement with the two possible TBA equations of
section 4.2, the two cases are to be treated with different partitions.

\subsection{Examples}

\subsubsection{The Ising Model}

The Ising model is the first example in the series of unitary minimal
models, i.e. ${\cal M}(3,4)$ with $c=1/2$. It has only three different
sectors. The anyonic $E_{8}$ related characters read 
\begin{equation}
\chi _{1,m}^{3,4}(q)=q^{h_{1,m}-\frac{1}{48}}\sum\limits_{\vec{k}}\frac{q^{%
\vec{B}_{1,m}^{3,4}\cdot \vec{k}}}{\left( q^{16}\right) _{k_{1}}\ldots
\left( q^{16}\right) _{k_{8}}}  \label{Isin}
\end{equation}
with 
\begin{equation}
^{\vec{B}_{1,1}^{3,4}=\{2,3,4,5,11,12,13,14\}}\quad \quad ^{\vec{B}%
_{1,2}^{3,4}=\{1,3,5,7,9,11,13,15\}}\quad \quad \,^{\vec{B}%
_{1,3}^{3,4}=\{1,4,6,7,9,10,12,15\}}\quad .
\end{equation}
Notice that $2\vec{B}_{1,3}^{3,4}-\vec{1}$ is precisely the set of exponents
of $E_{8}$. Moreover, one  observes that  all vectors $\vec{B}$ possess  
similar properties as 
the exponents, namely $B_i+B_{r+1-i}=b$. 
This indicates that in a similar way as the
exponents are related to the eigenvalues of the Coxeter transformation,
these values are also related to the eigenvalues of a real matrix with order 
$b$. One can therefore conjecture that the characters of the form (\ref
{anychar}) also possess some formulation involving the Weyl group.

Similarly as the fermionic realizations possess formulations in terms of
different Lie algebras, the anyonic counterparts also enjoy this property.
From (\ref{Isin}) we obtain easily the following anyonic type characters 
related to $A_1$ 
\begin{equation}
\chi _{1,2}^{3,4}(q)=q^{\frac{1}{24}}\sum\limits_{k}\frac{q^{k}}{\left(
q^{2}\right) _{k}}\,.
\end{equation}
Notice that for both cases $b=\frac{h}{2}+1,$ with $h$ being the Coxeter
number.

\subsubsection{The Tricritical Ising Model}

The next model in the series of unitary minimal models is the tricritical
Ising model ${\cal M}(4,5)$ with $c=7/10$. Similar to the fermionic 
representation the anyonic version is related to $E_{7}$ 
\begin{equation}
\chi _{2,m}^{4,5}(q)=q^{h_{2,m}-\frac{7}{240}}\sum\limits_{\vec{k}}\frac{q^{%
\vec{B}_{2,m}^{4,5}\cdot \vec{k}}}{\left( q^{10}\right) _{k_{1}}\ldots
\left( q^{10}\right) _{k_{7}}}
\end{equation}
with 
\begin{equation}
\vec{B}_{2,1}^{4,5}=\{1,3,4,5,6,7,9\}\quad ,\quad \vec{B}_{2,2}^{4,5}=%
\{1,2,3,5,7,8,9\}
\end{equation}
Once more we observe the feature that $2\vec{B}_{2,1}^{4,5}-\vec{1}$ is
precisely the set of exponents of $E_{7}$ and $b=\frac{h}{2}+1$.

As  was already observed in \cite{Rocha}, the characters
$\chi _{2,m}^{4,5}$ admit a product representation which, in our
 notations, corresponds to formulae of the type (\ref{TYPE2}). For instance,
\begin{equation}
 \chi _{2,1}^{4,5}(q) = 
 q^{\frac{49}{120}} \, \frac{\ln_{q^5}(\tau+ \frac 12) 
 \ln_{q^5}(4\tau+\frac 12) \ln_{q^5}(5\tau+ \frac 12) }
 { \ln_{q^5}(2\tau)\ln_{q^5}(3\tau)} \ .
\end{equation}
Rewriting this expression with the help of (\ref{notinv}) and (\ref{inv}),
we obtain
\begin{equation}
 \chi _{2,1}^{4,5}(q) = 
 q^{\frac{49}{120}} \sum_{\vec{k}} 
 \frac{q^{\frac 52 (k_1^2+ k_2^2 + k_3^2) + \vec{k}\cdot\vec{B}_{2,1}^{4,5}} }
 {(q^5)_{k_1}\ldots(q^5)_{k_5}} \quad
{\rm with}\quad
 \vec{B}_{2,1}^{4,5} = \{ 
  -\frac 32 , \frac 32 , \frac 52 , 2 , 3  \} \ .
\end{equation}
This provides an example of an anyonic type character involving
both types of anyonic quasi-particles (see the discussion in 
subsection 4.2). The formula (\ref{ceffe}) now acquires the form:
$c=\frac{6}{5\pi ^{2}} ( 3L(1/2) + 2L(1) ) = 7/10$.

\subsubsection{The Yang-Lee Model}

The Yang-Lee model ${\cal M}(2,5)$ (with $c=-22/5$)  
serves as the simplest example of a model
which is non-unitary. Its anyonic characters read 
\begin{equation}
\chi _{1,m}^{2,5}(q)=q^{h_{1,m}+\frac{11}{60}}\sum\limits_{\vec{k}}\frac{q^{%
\vec{B}_{1,m}^{2,5}\cdot \vec{k}}}{\left( q^{5}\right) _{k_{1}}\left(
q^{5}\right) _{k_{2}}}
\end{equation}
with 
\begin{equation}
\vec{B}_{1,1}^{2,5}=\{2,3\}\quad ,\quad \vec{B}_{1,2}^{2,5}=\{1,4\}\,.
\end{equation}
Notice that precisely these characters correspond to the original identities
of Rogers-Ramanujan and Schur \cite{RSR}. For this case we also illustrate
the working of formula (\ref{quasis}) and present an explicit construction
of the quasi-particle states of the lowest levels in table 1.

\begin{table}[h]
\begin{tabular}{|l|l|llll|}
\hline\hline 
l & d & $ \vec{p}(\vec{m})  $ & & & \\ [0.7mm] \hline\hline
2 & 1 & $  p_{1}^{1}(1,0)=2  $ & & & \\ [0.7mm] \hline\hline
3 & 1 & $  p_{2}^{1}(0,1)=3  $ & & & \\ [0.7mm] \hline\hline
4 & 1 & $ p_{1}^{1}(2,0)=-\frac{1}{2}$ & 
        $p_{1}^{2}(2,0)=\frac{9}{2}$ & & \\ [0.7mm] \hline\hline
5 & 1 & $ p_{1}^{1}(1,1)=\frac{9}{2} $ &  
        $p_{1}^{2}(1,1)=\frac{1}{2}  $ & & \\ [0.7mm] \hline\hline
6 & 2 & $p_{2}^{1}(0,2)=\frac{1}{2} $ & 
        $p_{2}^{2}(0,2)=\frac{11}{2} $ & & \\ [0.5mm]
 & &    $p_{1}^{1}(3,0)=-3 $ & 
        $p_{1}^{2}(3,0)=2 $ & 
        $p_{1}^{3}(3,0)=7$ & \\ [0.7mm] \hline\hline
7 & 2 & $p_{1}^{1}(2,1)=7 $ & 
        $p_{1}^{2}(2,1)=2 $ & 
        $p_{2}^{1}(2,1)=-2 $ & \\ [0.5mm]
 & &    $p_{1}^{1}(2,1)=-3 $ & 
        $p_{1}^{2}(2,1)=7 $ & 
        $p_{2}^{1}(2,1)=3 $ & \\ [0.7mm] \hline\hline
8 & 3 & $p_{1}^{1}(1,2)=-3 $ & 
        $p_{2}^{1}(1,2)=3 $ & 
        $p_{2}^{2}(1,2)=8 $ &  \\ [0.5mm]
 & &    $p_{1}^{1}(1,2)=2 $ & 
        $p_{2}^{1}(1,2)=-2 $ & 
        $p_{2}^{2}(1,2)=8 $ &  \\ [0.5mm]
 & &    $p_{1}^{1}(4,0)=-\frac{11}{2} $ & 
        $p_{1}^{2}(4,0)=-\frac{1}{2} $ &
        $ p_{1}^{3}(4,0)=\frac{9}{2} $ & 
        $p_{1}^{4}(4,0)=\frac{19}{2} $ \\ [0.7mm] \hline\hline
9 & 3 & $p_{2}^{1}(0,3)=-2 $ & 
        $p_{2}^{2}(0,3)=3 $ & 
        $p_{2}^{3}(0,3)=8 $ &  \\ [0.5mm]
 & &    $p_{1}^{1}(3,1)=-\frac{1}{2} $ & 
        $p_{1}^{2}(3,1)=\frac{9}{2} $ & 
        $p_{1}^{3}(3,1)=\frac{19}{2} $ & 
        $p_{2}^{1}(3,1)=-\frac{9}{2} $ \\ [0.7mm]
 & &    $p_{1}^{1}(3,1)=-\frac{11}{2} $ & 
        $p_{1}^{2}(3,1)=\frac{9}{2} $ & 
        $p_{1}^{3}(3,1)=\frac{19}{2} $ & 
        $p_{2}^{1}(3,1)=\frac{1}{2}$ \\ [0.7mm] \hline\hline
\end{tabular}
\caption{Anyonic quasi-particle states for $\chi _{1,1}^{2,5}$}
\end{table}

\section{Conclusions}

We conclude by summarizing our main results and comment on further problems.
We have obtained character formulae of the Virasoro algebras which are
related to simply laced Lie algebras and admit an anyonic quasi-particle
interpretation. As the main technical tool we exploited the fact that some
characters factorize in terms of quantum dilogarithms. Performing the
classical limit we also conjectured an anyonic version of the thermodynamic
Bethe ansatz.

In order to support the picture we have been presenting, it is of course
highly desirable to derive \cite{ABAF} the TBA-equations from proper
thermodynamic principles in the spirit of Yang and Yang \cite{Yang}. We
would like to emphasize once more that the proposed anyonic TBA-equations
are by far simpler than the usual fermionic or bosonic versions. The problem
of solving the set of non-linearly coupled equations to find the
pseudo-energies does not occur at all in the anyonic formulation. This
simplification is contrary to most situations in which exotic statistics is
implemented. The explicit construction of anyonic creation and annihilation
operators in momentum as well as in real space is far more complicated than
the construction of their bosonic and fermionic counterparts. Also the
scattering matrix, despite the fact that one may write it down easily, is an
object which is difficult to handle due to its complicated analytical
structure. 

What the conformal side concerns, the anyonic picture of the Virasoro
characters is also simpler. For instance, in comparison with the fermionic
realizations, the relatively complicated restrictions on the sums have
entirely vanished. It would be interesting to obtain the lattice version of
the states (equivalently to \cite{WP,Berk}) constructed 
in the previous section as well as
their explicit representations in the spirit of 
\cite{FeiginSto,Bou}. An investigation of how
the integrals of motion act on these states, i.e. to find their eigenvalues
like in \cite{BF}, will certainly help in this direction.

{\bf Acknowledgment: } We would like to thank H.M. Babujian, A.A. Belavin, 
W. Eholzer, A. Foerster, M. Karowski, K.P. Kokhas, M. Martins, 
J. Mund, A. Recknagel, 
R. Schrader and B.
Schroer for useful comments and discussions. A.B. is grateful to the members
of the Institut f\"{u}r Theoretische Physik, FU-Berlin for their hospitality.

\small
\setlength{\baselineskip}{12pt}

\end{document}